\documentstyle[sprocl,epsfig]{article}

\def\ra{\rightarrow}

\def\be{\begin{equation}}
\def\ee{\end{equation}}
\def\bea{\begin{eqnarray}}
\def\eea{\end{eqnarray}}

\begin{document}

\title{STRONG INTERACTION EFFECTS IN SCALAR TOP PRODUCTION}

\author{ M. DREES }

\address{Lab de Physique Mathematique, Univ.\ de Montpellier II,
France}

\author{ O.J.P. \'EBOLI }

\address{Inst.\ de F\'{\i}sica T\'eorica, Univ.\ Estadual Paulista, S\~ao
Paulo, Brazil}

\maketitle

\abstracts{
We discuss effects of fragmentation and hard gluon radiation on the
signal for the pair production of the lighter scalar top eigenstate
$\tilde{t}_1$ at $e^+e^-$ colliders. The main emphasis is on scenarios
with small stop--LSP mass splitting, where strong interaction effects
can considerably modify kinematical properties of the final state.}
  
The search for supersymmetry is on the ``to do'' list of all experiments
at currently operating or planned high energy particle colliders. It
is commonly believed that hadron colliders are better suited for the
search for strongly interacting sparticles (gluinos and squarks) than
lepton colliders are. Hadron colliders indeed offer larger cross sections
for producing such sparticles. However, experiments at hadron colliders
also suffer from much larger backgrounds; severe cuts are then needed
to extract a signal for sparticle production \cite{dpf}. As a result,
the production of sparticles will become undetectable if their decays
only release a small amount of visible energy.

This might well be the case for the lighter stop mass eigenstate
$\tilde{t}_1$. It is expected to be significantly lighter than the
other squarks, due to the renormalization group running of soft breaking
stop masses as well as mixing between $\tilde{t}_L$ and $\tilde{t}_R$
current eigenstates \cite{rudaz}. In fact, $m_{\tilde{t}_1}$ could be
quite close to the mass of the LSP, usually assumed to be the lightest
neutralino $\tilde{\chi}_1^0$. In this case $\tilde{t}_1$ becomes
quite long--lived, decaying primarily into $q \tilde{\chi}_1^0 \ (q=u,c)$
through $\tilde{t} - \tilde{q}$ mixing \cite{ken}. Since flavor
mixing in the squark sector has to be small in order to satisfy FCNC
constraints, and might in fact occur only at the one--loop level, the
lifetime of $\tilde{t}_1$ will then exceed the hadronization time scale,
i.e. $\tilde{t}_1$ will fragment into a (spin$-1/2$) stop ``meson''
before it decays \cite{ken}.

Implementing this fragmentation in an event generator is not entirely
straightforward. The reason is that fragmentation of a massive
on--shell squark into a hadron with equal or greater mass violates the
conservation of energy and/or 3--momentum. In ref.\cite{de1} we have
developed two algorithms to solve this problem. Briefly, method I
allows energy conservation to be violated in the fragmentation step,
but takes care to conserve 3--momentum; overall energy conservation is
then restored by rescaling all 3--momenta by a common factor, which is
calculated numerically by iteration. In the second method, the stop
squarks are created off--shell; the fragmentation step can then
conserve both energy and 3--momentum, being formally equivalent to the
(collinear) decay of a (virtual) squark with mass $m_*$ into a
``meson'' with mass $m_{\tilde{t}_M} < m_*$ and a massless (color
triplet) ``fragmentation jet''. The distribution in $m_*$ is chosen
such that the energy distribution of the stop ``meson'' resembles that
predicted by the assumed fragmentation function.

A second ambiguity is related to the choice of the fragmentation
variable $x$. For fixed functional form of the fragmentation function
$D_{\tilde t}(x)$, defining $x$ as the ratio of ``meson'' and squark
energies (in the lab frame) will give a much softer stop ``meson''
energy spectrum than defining $x$ as the ratio of ``meson'' and squark 
3--momenta. Fortunately we found \cite{de1} that much, although not all,
of the differences between the stop ``meson'' energy spectra predicted 
by our two different algorithms for stop fragmentation, or using
different definitions of the fragmentation variable $x$, can be
absorbed in a re--definition of the single parameter $\epsilon$ of the
Peterson et al. fragmentation function \cite{peter}.

\begin{figure}[htb]

\vspace*{-3cm}
{}

\centerline{\epsfig{file=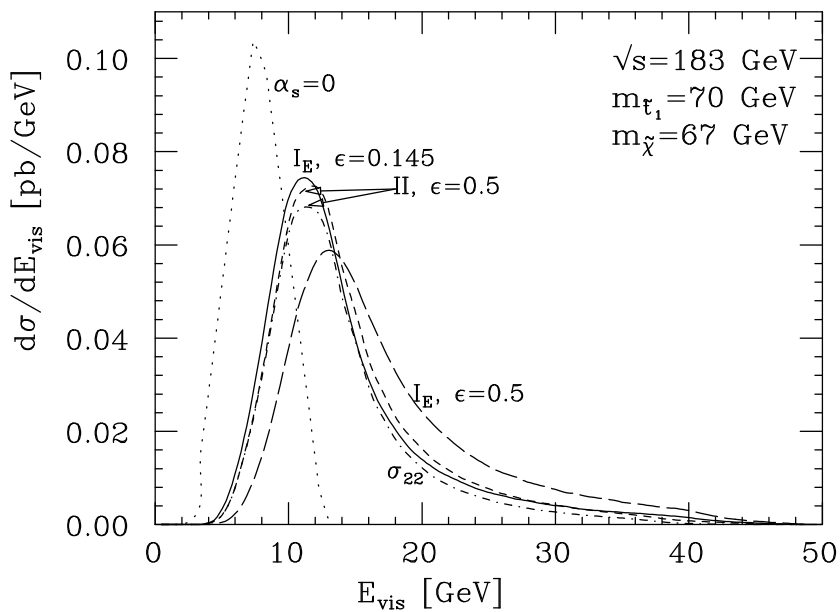,height=13cm}}

\caption
{The total visible energy in $e^+e^- \ra \tilde{t}_1 \tilde{t}_1^*
(g)$ events at LEP--183. The dotted curve has been obtained by
switching off both perturbative and non--perturbative strong
interaction effects, while the dot--dashed curve includes
non--perturbative $\tilde{t}_1$ fragmentation but no hard gluon
radiation.  The other curves include both perturbative and
non--perturbative effects. The labels refer to the fragmentation
scheme used, and give the value of the fragmentation parameter
$\epsilon$ in units of GeV$^2$.}
\end{figure}

Additional contributions to the visible energy can come from
perturbative QCD processes. In particular, the $\tilde{t}_1$ pair can
be produced in association with a hard gluon \cite{glu,de1}. As shown
in Fig.~1, this perturbative contribution to the visible energy is
actually smaller than the non--perturbative one, at least for beam
energies not far above the $\tilde{t}_1$ pair threshold. It
nevertheless plays an important role in searches for $\tilde{t}_1$
production in scenarios with very small $\tilde{t}_1 -
\tilde{\chi}_1^0$ mass difference $\Delta m$. In the absence of hard
gluon radiation the final state then consists essentially of the two
fragmentation jets, which are back--to--back in the transverse
plane. Such final states suffer very large backgrounds from
two--photon processes. Experimental searches for $\tilde{t}_1$
production at LEP \cite{search} therefore require the final state to
have an acoplanarity angle significantly smaller than
$180^\circ$. This becomes possible in the presence of hard gluon
radiation even if $\Delta m \ra 0$. As shown in Fig.~2, $\tilde{t}_1
\tilde{t}_1^* g$ events should therefore allow to extend LEP searches
to the case $\Delta m \ra 0$, although the reach in $m_{\tilde{t}_1}$
will be smaller than for larger values of $\Delta m$. 

\begin{figure}[htb]

\vspace*{-3cm}

\centerline{\epsfig{file=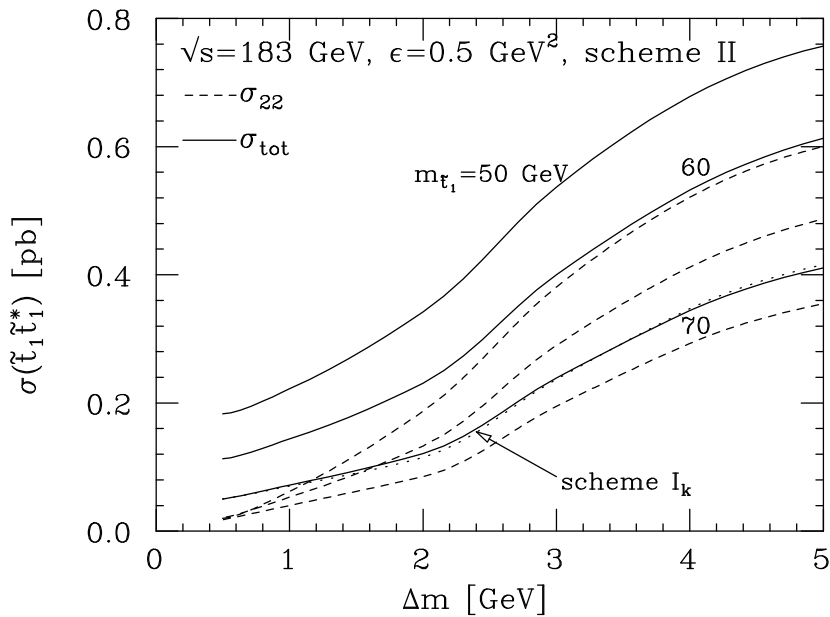,height=13cm}}

\caption
{The total inclusive $\tilde{t}_1 \tilde{t}_1^*$ cross section at
LEP--183 after cuts similar to those used by the OPAL collaboration
have been applied. The solid curves show the full result, while the
dashed curves have been computed by switching off hard gluon
radiation. All curves are for fragmentation scheme II, except
the dotted one, which is for scheme I$_k$.}

\end{figure}

Analogous results should hold for $\tilde{t}_1$ searches at future
linear colliders. However, there the size of the two--photon background,
and hence the potential for $\tilde{t}_1$ searches in scenarios
with small $\Delta m$, depends quite sensitively on details of the
design of both the accelerator and the detector. The former determines
the flux of beamstrahlung photons, which can easily dominate the total
effective photon flux in the relevant energy range of a few GeV. The 
latter will e.g. determine the minimal angle at which outgoing 
$e^\pm$ can be vetoed, and hence the maximal missing transverse
momentum in two--photon events. We therefore do not attempt to repeat
an analysis like that shown in Fig.~2 for higher energies.

Finally, we remark that QCD corrections to $\tilde{t}_1$ decay were
not included in our results. An accurate treatment of these decays is
not very important if $\Delta m$ is small, since kinematical
constraints then imply that the contribution to the visible energy
from these decays is in any case small. This is fortunate, since for
$\Delta m \leq 2$ GeV a perturbative treatment of these decays becomes
altogether unreliable. On the other hand, QCD effects are expected to
be important also for the interpretation of precision measurements of
squark properties, in particular for the determination of their
masses, even if $\Delta m$ is large; the possibility to perform such
precision studies is another major selling point of $e^+e^-$
colliders. In this case a careful treatment of squark decays, including
QCD corrections, becomes important. An investigation of these effects
is now in progress \cite{degk}.


\vskip 8pt

\noindent{\bf Acknowledgements}

\vskip 8pt

This work was supported in part by Conselho Nacional de Desenvolvimento
Cient\'{\i}fico e Tecnol\'ogico (CNPq), by Funda\c{c}\~ao de Amparo \`a
Pesquisa do Estado de S\~ao Paulo (FAPESP), and by Programa de Apoio a
N\'ucleos de Excel\^encia (PRONEX).

\clearpage

\end{document}